\documentclass[preprint, 5p]{elsarticle}
\pdfoutput=1
\usepackage{epsfig}
\usepackage{amssymb}
\usepackage{amsmath}
\usepackage[utf8]{inputenc}
\journal{Physical Review Letters}
\usepackage{subfigure}
\usepackage{color}
\usepackage{lineno}
\usepackage{float}
\usepackage{bm}
\usepackage{titlesec}
\usepackage{graphicx}
\usepackage{mwe}
\usepackage{subfig}
\usepackage{hyphenat}
\titleformat{\paragraph}[runin]
{\bfseries\scshape}{\theparagraph}{1em}{}
\setcitestyle{square}

\begin{document}
\sloppy
\begin{frontmatter}
\date{\today}

\title{On ANITA's sensitivity to long-lived, charged massive particles}


\author[OSU]{Amy Connolly}
\author[OSU]{Patrick Allison}
\author[OSU]{Oindree Banerjee}
\address[OSU] {Dept. of Physics, Center for Cosmology and AstroParticle Physics, Ohio State Univ., Columbus, OH 43210.}

\begin{abstract}

We propose that the Antarctic Impulsive Transient Antenna (ANITA) can serve as a detector for long-lived, charged particles, through its measurement of extensive air showers from secondary leptons. To test this on an example model, we simulate the production of staus inside the earth from interactions between ultra-high energy neutrinos and nuclei. 
 We propose that results
 of ANITA searches for upgoing air showers can be interpreted in terms of constraints on long-lived, charged massive particles (CHAMPs) and consider a supersymmetric partner of the tau lepton, the stau, as an example of such a particle.  Exploring the parameter space in stau mass and lifetimes, we find that the stau properties that lead to an
 observable signal in ANITA are highly energy dependent.
At $10^{18.5}$~eV, we 
 find that the best constraints on the product of
 the neutrino flux and the stau production 
 cross section would
 be placed near $m_{\tilde{\tau}}=$1~TeV and $\tau_{\tilde{\tau}}=$10~ns.
 Thus ANITA could be sensitive to new physics in a region of 
 parameter space that is unconstrained by 
 experiments at the
 Large Hadron Collider.

\end{abstract}


\end{frontmatter}

\section{Introduction}
\label{motivation}

ANITA is primarily designed as an ultra-high energy (UHE) neutrino discovery experiment, operating under 
NASA's long-duration balloon program~\cite{design}.
ANITA searches for broadband, impulsive signals at radio frequencies expected from Askaryan radiation produced
from neutrino-induced cascades in the ice.

ANITA also serves as a UHE cosmic ray detector~\cite{hoover}. Extensive air showers (EAS) due to cosmic rays produce synchotron and Askaryan emission in
the local, nearly-vertical geomagnetic field, observed by 
ANITA as
horizontally-polarized (HPol) pulses, in contrast to
the neutrino signature described above where events are
expected to be predominantly vertically-polarized (VPol). ANITA detects these air-shower-induced 
pulses mainly after their reflection off of the ice, where the signals acquire a polarity opposite from those observed directly. Events observed directly are typically seen at elevation angles above the horizon as seen by ANITA, which is 6.5 degrees down from the horizontal. 

Long-lived, charged massive particles (CHAMPs) produced in the
earth could be detected by ANITA if they 
ultimately lead to an observable EAS.
Since EAS signals arriving from elevation angles steeper than 6.5 degrees are expected to have obtained an opposite polarity from a reflection, upward-going
signals
with a non-inverted polarity could be a signature
for an association with a CHAMP.

Models that lead to upward-going
air showers observable by ANITA have heightened 
interest with ANITA's reporting of two upward-going, unusual events~\cite{me1,me2}.
Cherry \textit{et al.}~\cite{2018_sterile_nu} have proposed a sterile neutrino origin for these events, and Anchordoqui \textit{et al.}~\cite{2018_cpt_symm} extends this work to accommodate a massive dark matter candidate (480 PeV right-handed neutrino), trapped in the earth, in place of the sterile neutrino. The dark matter particle could decay into a Higgs and a light Majorana neutrino, the latter of which would produce a lepton in the crust causing the observed air shower. Both of these possibilities may have significant problems producing the observed arrival directions at ANITA without overproducing at other angles.  Anchordoqui \textit{et al.} addressed this by evoking an atypical dark matter density distribution in the Earth.

We find that the CHAMPs investigated here would suffer a similar problem if they were to be posed as an explanation
of the unusual events reported by ANITA;
a preference for steeply upward-going events comes in 
a region of parameter space with far-less-than-optimal
event rates that are not expected to be detectable. 
Still, 
we report on the sensitivity of ANITA to CHAMPs coming from any angle as a
new search channel for the experiment.

\section{CHAMPs and neutrino telescopes}

Long-lived, charged particles have long
appeared in  
extensions to the Standard Model (SM) of particle physics.  For our investigation we consider a long-lived charged slepton that arise in some supersymmetric models.

Supersymmetry (SUSY) 
offers a symmetry between bosons and fermions that could offer a solution to the
fine-tuning problem that plagues the 
SM, as well as a possible unification of couplings at high energies.  Requiring
R-parity conservation requires that the Lightest SUSY Particle (LSP) be stable, and depending on the scale, this can be the gravitino. In this case, the Next to Lightest SUSY Particle (NLSP) can be a charged slepton, specifically, the right-handed stau. Moreover, the decay of the slepton can be
kinematically suppressed, and thus its lifetime
long, if the difference between the masses of the LSP and NLSP is small.

Collider searches have placed lower bounds of about $450\,\textrm{GeV}$ on the mass of a long-lived charged slepton \cite{ATLAS:2014fka,Khachatryan:2016sfv,Abazov:2012ina,Chatrchyan:2013oca} for cases where the particle lives long enough to leave the detector. Direct stau production has been ruled out for masses below about $300\,\textrm{GeV}$ \cite{ATLAS:2014fka}. 
These bounds are typically based on some model
assumptions.

Neutrino observatories have been previously proposed as detectors of long-lived, supersymmetric particles.
Albuquerque \textit{et al.}~\cite{Albuquerque:2003mi,Albuquerque:2007} and Ahlers \textit{et al.}~\cite{Ahlers:2006zm,staus_cr} point out that km$^3$-scale neutrino telescopes could inform the SUSY breaking scale, while complementing results from the Large Hadron Collider (LHC.)

In the studies in~\cite{Albuquerque:2003mi,Albuquerque:2007,Ahlers:2006zm}, UHE neutrinos 
produce stau pairs
through interactions 
with a nucleon in the earth.
The proposed signature for the observation of staus is the detection of two parallel charged tracks separated by about 100 meters 
predicting at least a few events per year for an optical Cherenkov detector such as IceCube. In~\cite{staus_cr}, they considered cosmic rays rather than neutrinos producing the 
staus, and \cite{Ando:2007ds} considered both cases.  It is expected that 
neutrinos would produce the dominant flux for most typical supersymmetric models, since cosmic rays primarily interact via Standard Model strong interactions \cite{Ando:2007ds}.

The requirement of a stau pair with nearby parallel tracks in the searches proposed in ~\cite{Albuquerque:2003mi,Albuquerque:2007, Ahlers:2006zm,staus_cr} is motivated by the obstacle of distinguishing between high energy stau and low energy muon tracks as IceCube is particularly susceptible to the latter.
However, this strongly limits the detection ability at extremely high energies, where the effective rate can be suppressed by a factor of 1000 or more \cite{Kersten:2006yq}. In addition, in the model where only one of the supersymmetric particles produced is a stau, the only detection channel open for IceCube would be where the tau produced by the stau decay interacted inside the IceCube volume directly.
ANITA, on the other hand, does not suffer from the challenges of a muon-dimuon background, so an argument for a single stau detection is conceivable.

\section{A preference for shallow upward-going events}

\begin{figure}
\centerline{\includegraphics[width=8.0cm]{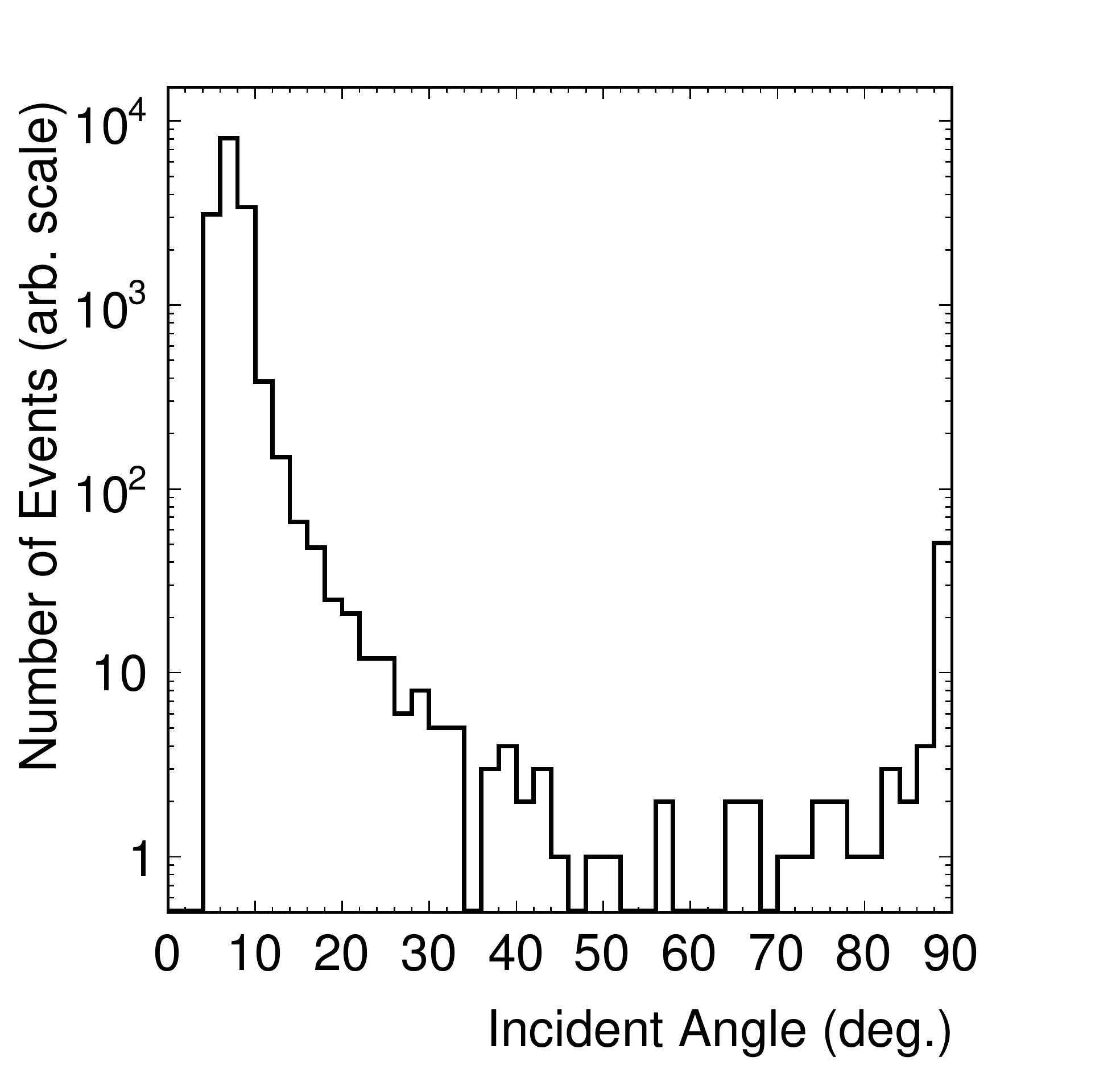}}
\caption{\label{fig:incident_angles} This figure shows
the distribution of incident angles below
horizontal as seen by an observer at the position of the ANITA payload, based
purely on geometric considerations.  Here
we have taken 
an isotropic
flux of particles incident uniformly across the
surface of the earth {\it in the absence of any absorption}.
}
\end{figure}

Based on purely geometric considerations, and assuming an isotropic incident flux with
neutrino interactions turned off,
Fig.~\ref{fig:incident_angles} shows a strong preference for shallow trajectories as seen by ANITA. 
This is because a particle incident with a  
steep trajectory from the far side of 
the earth
sees ANITA's 700~km-radius
ice target as covering a small solid angle,
while a neutrino arriving at a more shallow angle sees the ice in ANITA's horizon 
at a shorter distance.  Coming from below, however, ANITA's target looks like a circle
while from a shallow angle it is viewed at 
an angle, lowering the probability of it being intersected by the particle trajectory.

Roughly, for an isotropic incident flux,
the ratio of probabilities that
an event arriving at a steep angle emerges within 
ANITA's horizon compared to the probability that
one at a shallow angle does can then be estimated as:
\begin{equation}
\dfrac{P(\theta_{\rm{steep}})}{P(\theta_{\rm{shallow}})}\approx \dfrac{c^2_{\rm{shallow}}}{c^2_{\rm{steep}}} \dfrac{ \sin{\theta_{\rm{steep}}}}{\sin{\theta_{\rm{shallow}}}}
\end{equation}
where the $c$'s are the chord lengths traversed by
each type of trajectory.  If we consider $\theta_{\rm{shallow}}=6.5^{\circ}$ and $\theta_{\rm{steep}}=35^{\circ}$, then taking the
earth radius $R_{\rm{earth}}=6.36\times10^6$~m,
then $c_{\rm{shallow}}=460$~km and $c_{\rm{steep}}=7200$~km, giving:
\begin{equation}
\dfrac{P(\theta_{\rm{steep}})}{P(\theta_{\rm{shallow}})}\approx 2\%. 
\end{equation}

As shown in Fig.~{\ref{fig:incident_angles}}, using a toy Monte Carlo we find the distribution of incident angles for an isotropic flux of completely lossless particles, and find
that approximately 1.7\% of those
would be seen by ANITA to have incident angles 
greater than $27^{\circ}$.  However, this includes
the uptick near $90^{\circ}$ incident angles
where ANITA's ice target is seen straight on as a circle.  

If we only consider incident angles in the range $27-50^{\circ}$ they account for only
0.7\% of trajectories that emerge in ANITA's horizon, and again this does not include neutrino absorption in the earth that would only stronly suppress steep events.
We point out that this purely geometric argument means that any model
that is to explain a preference for steep incoming
trajectories over shallow ones, if assuming
an isotropic neutrino flux, must overcome this bias for shallow trajectories
by about a factor of $\sim 100$. We note however, that ANITA does consider 
point sources as the origin of
neutrinos leading to the upward-going events.

We note that the IceCube detector has reported
a high-energy, through-going track event with $2.6$~PeV deposited
energy \cite{2015ATel.7856....1S}, implying a neutrino flux at 
about $10$~PeV, but upward-going at $\sim10^{\circ}$.
As noted in \cite{2016arXiv160508781K}, a lack of events at shallower and
downward-going angles is interesting, and motivates a mechanism that might
preferentially produce steeply upward-going events. It is also worth noting that
the Pierre Auger Observatory, the only other similar-scale experiment in this
energy range, would likely not be able to observe similar steeply upward-going
events with their standard analysis (intended for earth-skimming neutrinos less than $5^{\circ}$ from the horizontal) \cite{Aab:2015kma}, since showers more than $\sim3^{\circ}$ below horizontal would not produce significant particle flux at
ground \cite{Zas:2005zz}.


\section{Modeling upward-going taus enabled by long-lived stau leptons}

\begin{figure}
\centerline{\includegraphics[width=6.0cm]{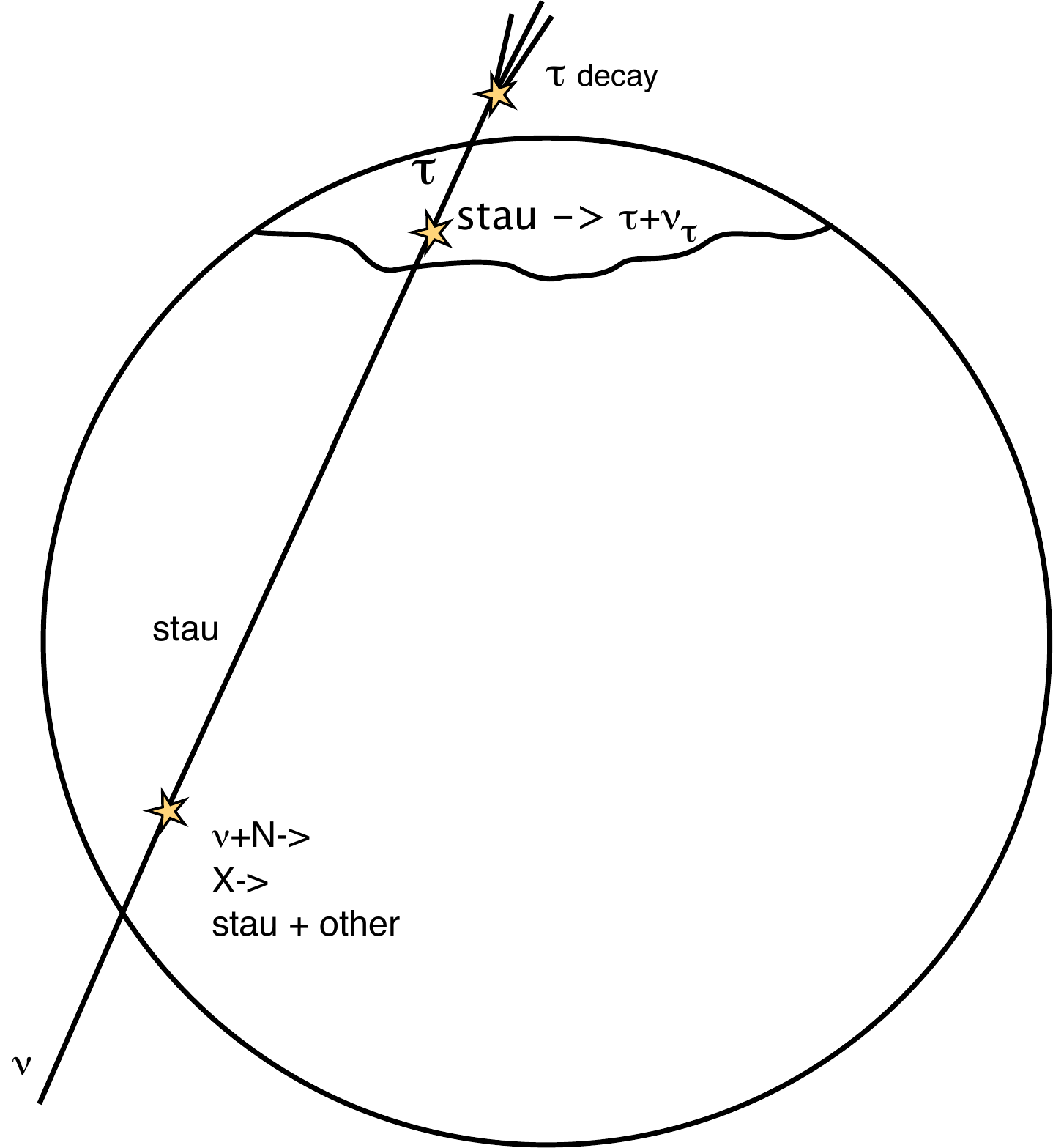}}
\caption{\label{fig:earth_picture} Sketch of
the signature being considered here.  Although
the figure shows one stau being produced in 
the $\nu N$ interaction, a stau pair could
be produced instead, doubling the probability of
detection.}
\end{figure}

We wish to estimate the probabilities for neutrinos at different
energies
to penetrate the earth at different incident angles as seen
by ANITA through the signature shown in Fig.~\ref{fig:earth_picture}, that is,
single stau production
where a tau emerges from the earth and decays
in air.
For this 
we utilize both a toy Monte Carlo and a separate
integration of the probability that a
neutrino will lead to an observable air shower for a 
given incident angle.
We note that stau pair production might also contribute,
as would events where a tau decays close enough to the 
surface for the shower to emerge, but we do not 
consider either of those at this stage.
We consider both an isotropic neutrino flux 
(arriving from 2$\pi$ above the surface at
any entrance point), and a single input angle
of incidence, and also model both flux spectra
and sets of monoenergetic neutrinos.  

Our simulation is kept relatively simple
at this stage to simplify understanding.
At this stage, we give the entire earth the density
of the mantle, and we propagate energy losses
of both the stau and the tau as described below.  At each interaction, the energy
of the decay product of interest is just half
of its parent's, so that the initial stau energy is $E^i_{\rm{stau}}=E_{\nu}/2$, the initial
tau energy is
$E^i_{\rm{tau}}=E^f_{\rm{stau}}/2$, and
the energy of the shower produced by the tau
is $E_{\rm{shower}}=E_{\rm{tau}}/2$.  We take the threshold
for detectability of a shower at ANITA to be $10^{17}$~eV
when at 100~km distance, taking the signal strength of any 
given shower to be proportional to energy and inversely proportional to distance.
A more thorough treatment of the energy distributions of decays products at each stage will be implemented at a stage beyond this
initial study.

In our Monte Carlo we use the SM $\nu N$
cross section from~\cite{Gandhi:1995tf,Connolly:2011vc} with no
SUSY enhancement.
In Fig.~2 of~\cite{Albuquerque:2003mi}, they show that the  $\nu N$
cross section from SM processes alone dominates by over an order of magnitude compared to any contribution to it involving the exchange of SUSY particles, for left-handed lepton mass m$_{\widetilde{\ell}_{L}}$ in the hundreds of~GeV, chargino mass m$_{\widetilde{W}}=250$~GeV, and squark mass m$_{\widetilde{q}}$=300 GeV.  
This qualitative conclusion is confirmed in~\cite{Huang:2006ie}
for 50-250~GeV stau masses. To produce the largest range possible and to simplify the parameter space, we consider the case of minimal charged-current interactions ($\sin\theta_{f}=0$).

We used the parametrization for the energy loss of the stau presented in
~\cite{Reno:2005si}, with
\begin{equation}
 -\left<\dfrac{dE_S}{dX}\right>
 =\alpha+\beta E
 \end{equation}
 \noindent with $\alpha = 2\times 10^{-3}$~GeV~cm$^2$g$^{-1}$, $E_0=10^3$~GeV, and: 
 \begin{equation}
 \beta = \beta_0 + \beta_1 \ln{\left( E/10^{10}~\rm{GeV}  \right) + \beta_2 \ln{\left( E_0/10^3 \rm{GeV}  \right)}} 
 \end{equation}
 with:
 
\centerline{ $\beta_0 = 5\times 10^{-9} ~ \rm{cm}^2 \rm{g}^{-1} ~(150~\rm{GeV}/m_{\rm{\widetilde{\tau}}} )$}
 \centerline{ $\beta_1 = 2.8\times 10^{-10}~  \rm{cm}^2 \rm{g}^{-1} ~(150~\rm{GeV}/m_{\rm{\widetilde{\tau}}} )$}
 \centerline{ $\beta_2 = 2\times 10^{-10} ~ \rm{cm}^2 \rm{g}^{-1} ~(150~\rm{GeV}/m_{\rm{\widetilde{\tau}}} )$}.

\section{A mechanism that can improve the odds for 
steep trajectories}
\begin{figure}
\centerline{\includegraphics[width=8.0cm]{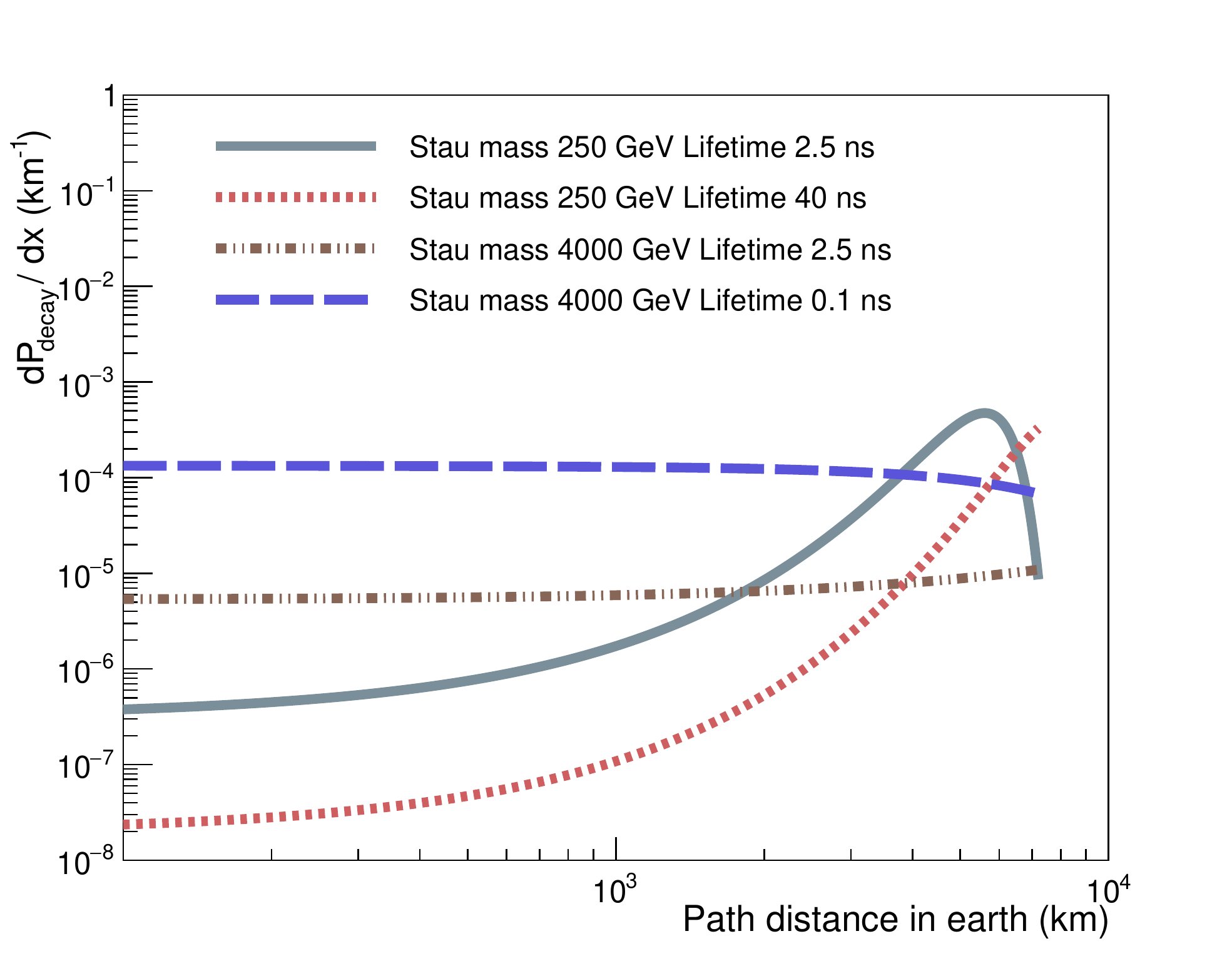}}
\caption{\label{fig:bragg-like} Differential probability of decay of a heavy, charged particle as a function of distance along its path through the earth.  Here, we consider different
combinations of lifetimes and masses that highlight different features that this distribution should take. We find that the parameter space that gives the best chance for ANITA to view this signature is one where the differential probability is still rising as the stau is nearing the surface, such as for $m_{\rm{stau}}=250$~GeV, $\tau_0=40$~ns. These curves are for a simulated stau with initial energy of $10^{22}$~eV. 
}
\end{figure}

A heavy, charged particle will undergo energy loss
along its journey through the earth due to ionization
and radiative losses.  A long-lived, energetic, 
heavy, charged particle, at production, will have a 
lifetime that is heavily dilated in the lab frame.
However, its lifetime in the lab frame will decrease 
as it loses energy along its path.  Therefore, for some combinations of the particle's mass, proper
lifetime and initial energy, if the particle decays
along its path in the earth, it will be most likely to decay close to its exit point.  If a long-lived stau
were to decay close enough to the surface of the earth,
then it could produce a tau that could decay in the air
and produce a shower that is visible by ANITA.

\begin{figure}[h]
\includegraphics[width=.48\linewidth]{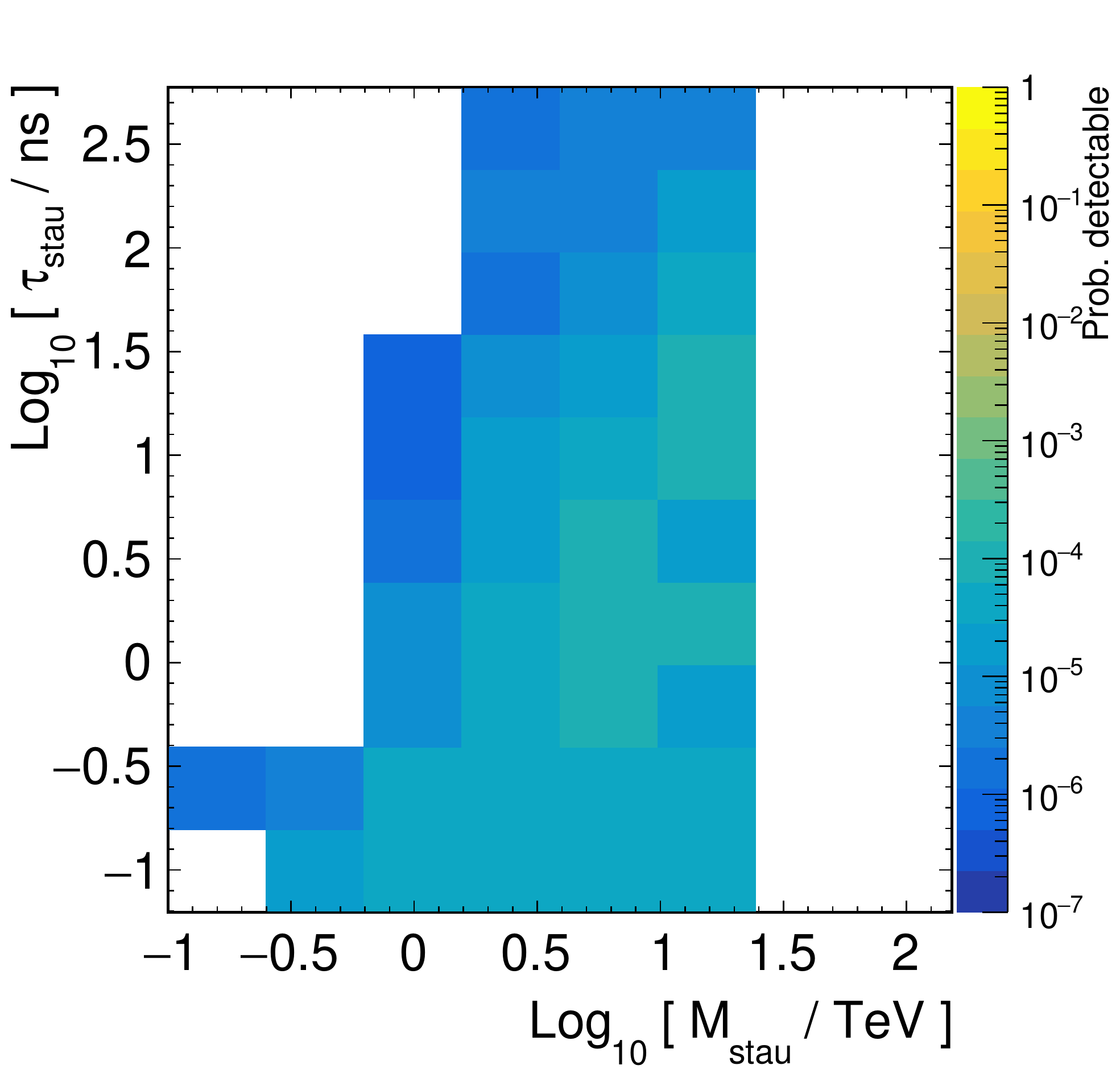}\hfill
\includegraphics[width=.48\linewidth]{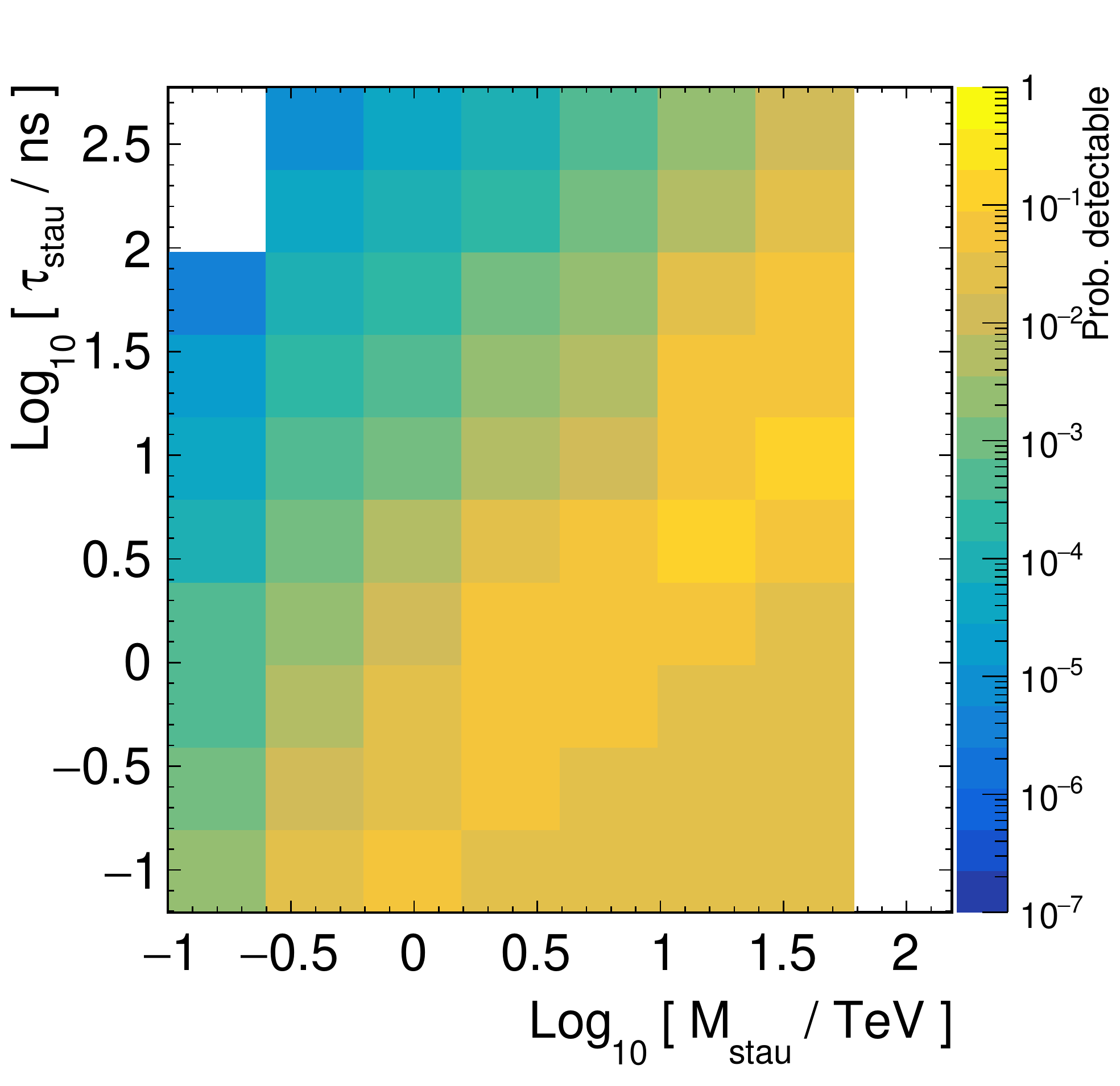}
\caption{\label{fig:probs}
Probability that a neutrino at a given energy incident at 
6.5$^{\circ}$ down from horizontal at
the balloon leads to a shower
in the air exceeding threshold through the signature considered
in this paper (see text), as a function of stau mass and lifetime. (Left) Neutrino energy $10^{18.5}$~eV and (right) 
neutrino energy 10$^{19}$~eV.}
\end{figure}
\begin{figure}[h]
\includegraphics[width=.48\linewidth]{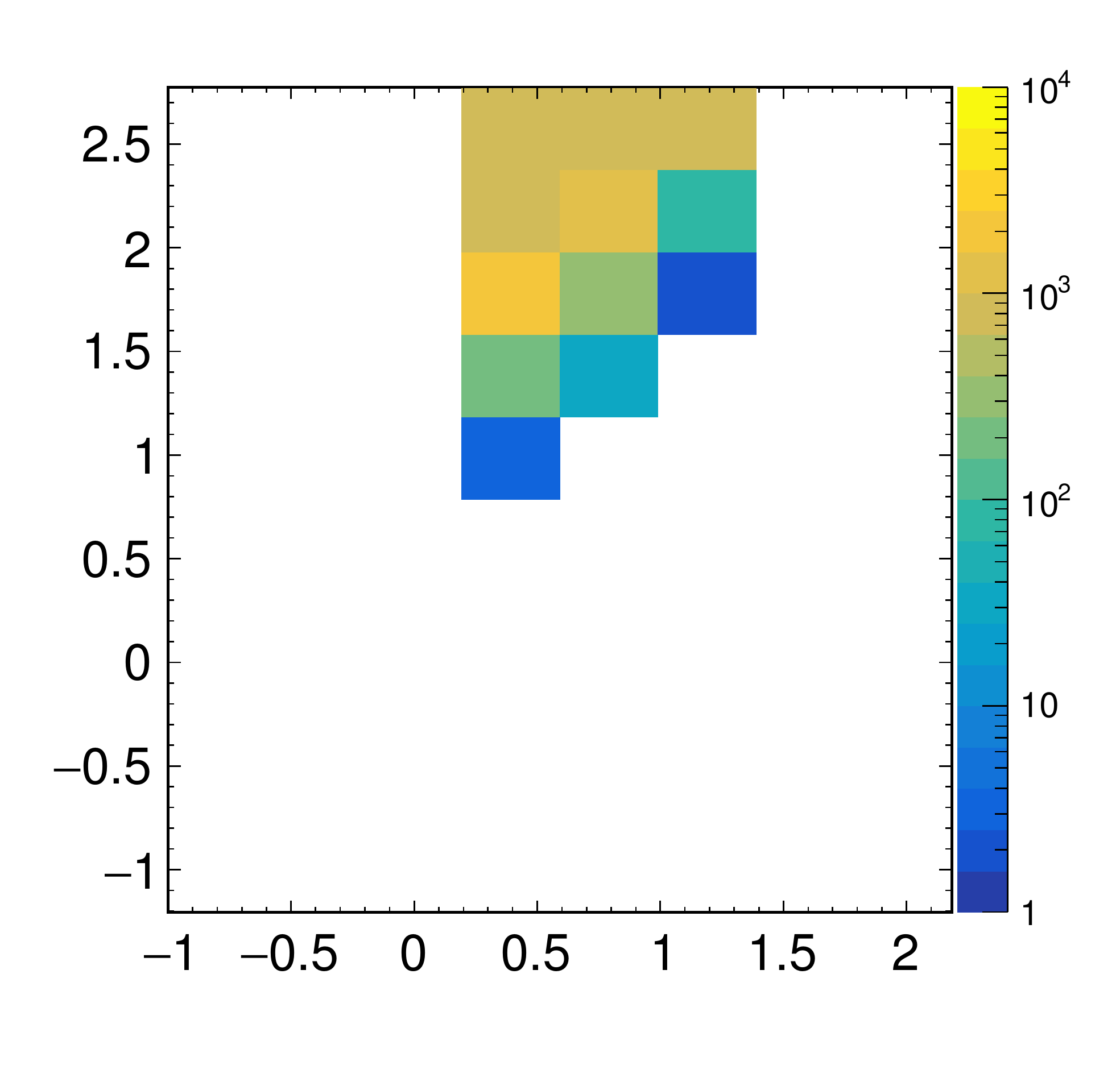}\hfill
\includegraphics[width=.48\linewidth]{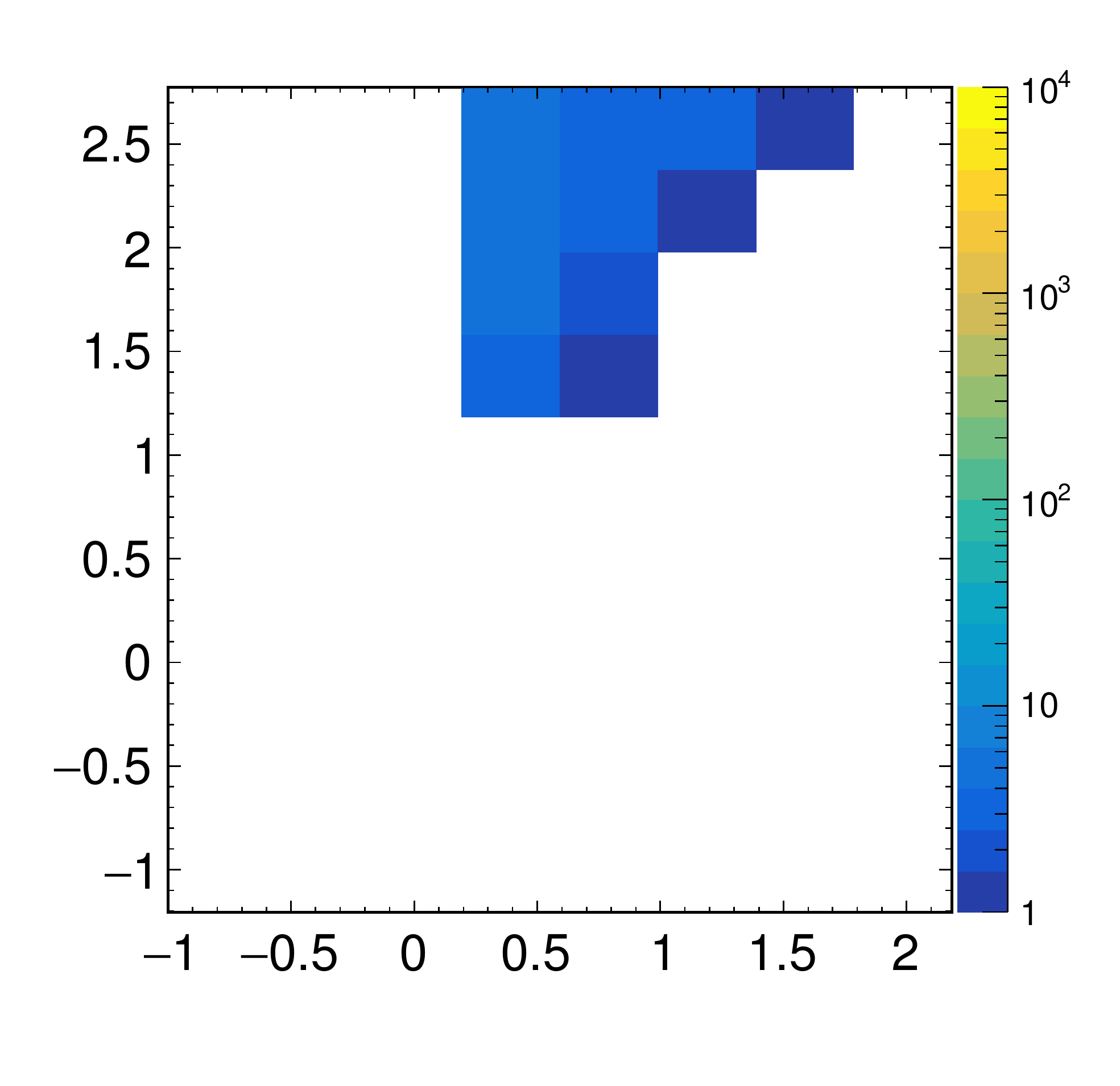}
\caption{\label{fig:ratios} 
Ratio of probabilities of a detectable tau-induced shower being produced at a 35$^{\circ}$ incident angle versus 
a 10$^{\circ}$ incident angle, as a function of stau mass and lifetime.
(Left) Neutrino energy $10^{18.5}$~eV and (right) 
neutrino energy 10$^{19}$~eV.
}

\end{figure}

We note that, as is shown in
~\cite{Reno:2005si}, the range of an UHE massive charged
particle is of order $10^4$~km, which is approximately
the chord length of a particle incident at $\sim 30^{\circ}$ angles (6000-7000~km for 27-35$^{\circ}$).
However, ANITA would not be able to observe showers
derived from decay products from staus that have come to rest.  That is because even
 UHE $\nu$-N interactions ($E_{\rm{cm}}=45$~TeV for
$E_{\nu}=10^{18}$~eV, increasing as $\sqrt{E_{\nu}}$),
cannot produce particles with masses that exceed
the shower energies of order $10^{17}$~eV that are
observable by ANITA.

We instead consider staus that decay before coming
to rest.
In Fig.~\ref{fig:bragg-like}, we show the
differential probability for a stau
to decay along its path through the earth
$dP_{\rm{decay}}/dx$ as a function of the distance $x$
from its entrance point.  All of these curves
are produced for an initial stau
energy of $10^{22}$~eV, varying
 the mass and proper lifetime of the particle
 in order to illustrate the important features.

For certain combinations of 
initial energy, mass and proper lifetime,
if the stau decays before leaving the earth, it is most
likely to decay near or at the end of its pass.  This
can be seen most dramatically for the grey, solid curve in Fig.~\ref{fig:bragg-like}
representing $m_{\tilde{\tau}}=250$~GeV and $\tau_0=2.5$~ns.  
This is reminiscent of the Bragg peak, where a particle leaves most of its energy 
at the energy of its track due to the steep rise
in $dE/dx\propto 1/\beta^2$ as $\beta$ decreases. 
However, in this case, the
peak is in differential decay probability, and it 
comes about due to the particle's decreasing
Lorentz factor
$\gamma$, and thus lifetime in the lab frame as it loses
energy along its path. This peak at a specific distance may give a boost to the probability
for ANITA to observe particles that have traversed
a significant distance through the earth.  
We attempt
to quantify this in the next section.

In Fig.~\ref{fig:probs}, we show the probability for a stau
produced from a neutrino that was incident at $6.5^{\circ}$ (from near the horizon)
to lead to 
 a shower above threshold as a function of 
the stau mass and lifetime for two different energies,
$10^{18.5}$~eV and $10^{19}$~eV.  The strong dependence
on neutrino energy is evident.  One can also see that
there is a diagonal region in mass-lifetime space that is 
most preferred for the signature to be detectable.  

In Fig.~\ref{fig:ratios}, we plot the ratio of the probabilities
that a signal is detectable if is were coming from
$35^{\circ}$ below horizontal compared to 
from $10^{\circ}$.
Where there is white space, there were no
events from one or both angles so that a ratio
could not be calculated.  Recall that ratios well in excess 
of 1000 would be needed in order to get a preference for
events at $35^{\circ}$ over ones at shallow angles, and we
do not find a region of parameter space that is able to overcome this factor.

\section{Conclusions}

We simulated the production of a stau inside the earth, from the interaction of a UHE neutrino and a nucleon to investigate the potential of ANITA as a detector for CHAMPs. 
Although we do not find that this signature explains an overall {\it preference}
for steep events ($>\approx 20^{\circ}$),
it is a mechanism that can help to skew 
the distribution of incident angles 
toward steeper angles on average.  

We note that the signature of upward-going showers is quite unique among
neutrino experiments in its geometry, 
propagating so steeply 
upward and moving from a more dense to less atmosphere.
This could lead to showers that begin in dense atmosphere and cease to develop where the air is thinner and thus have a lengthened shower, and thus more radio emission, which is seen to some degree
in experiments \cite{1475-7516-2016-09-024}. Further detailed modeling of the radio signal will
demonstrate whether the same effect is seen in ANITA.

We recommend that ANITA seek this signature and are optimistic that it will be able to 
place constraints at masses that are beyond
the reach of the LHC. We expect events
from this signature to be more likely to
come from shallow angles, while noting that 
a neutrino burst might be able to overcome the challenge of preferring steeper elevation angles.
We think that this would be an interesting
signature for other experiments searching for 
air showers from taus to investigate, such 
as BEACON, GRAND~\cite{Martineau-Huynh:2015hae} and TAROGE~\cite{Nam:2016cib}.

\section{Acknowledgements}
We would like to thank the National Science Foundation for CAREER Award 1255557 and also the Ohio Supercomputing Center.
We are grateful to Prof. John Beacom and Prof. Stephanie Wissel 
for helpful discussions.  We also thank Brian
Clark for his input.

\section{References}

\bibliographystyle{abbrv}
\refstepcounter{section}
\bibliography{staus}

\end{document}